# *Ab initio* investigation of hydrogen bonding and electronic structure of high-pressure phases of ice

Renjun Xu[a], Zhiming Liu, Yanming Ma, Tian Cui[b], Bingbing Liu, and Guangtian Zou

State Key Laboratory of Superhard Materials, Jilin University, Changchun, 130012, P. R. China

We report a detailed *ab initio* investigation on hydrogen bonding, geometry, electronic structure, and lattice dynamics of ice under a large high pressure range, including the ice X phase (55-380GPa), the previous theoretically proposed higher-pressure phase ice XIII[M] (Refs. 1-2) (380GPa), ice XV (a new structure we derived from ice XIIIM) (300–380GPa), as well as the ambient pressure low-temperature phase ice XI. Different from many other materials, the band gap of ice X is found to be increasing linearly with pressure from 55GPa up to 290GPa, the electronic density of states (DOS) shows that the valence bands have a tendency of red shift (move to lower energies) referring to the Fermi energy while the conduction bands have a blue shift (move to higher energies). This behavior is interpreted as the high pressure induced change of *s-p* charge transfers between hydrogen and oxygen. It is found that ice X exists in the pressure range from 75GPa to about 290GPa. Beyond 300GPa, a new hydrogen-bonding structure with 50% hydrogen atoms in symmetric positions in O-H-O bonds and the other half being asymmetric, ice XV, is identified. The physical mechanism for this broken symmetry in hydrogen bonding is revealed.

---

[a] Current address: Department of Physics, University of California, Davis, CA 95616, USA

[b] Author to whom correspondence should be addressed. Electronic address: cuitian@jlu.edu.cn

# I. INTRODUCTION

The high-pressure behavior of $H_2O$ is of fundamental importance to many processes in physics, chemistry, life science, and planetary physics.[3,4,5,6] Studies of water are also of general theoretical importance in giving insights into the common regularities of structure and polymorphism of various substances featuring fourfold coordination of molecules or atoms. In particular, it sheds light onto one of the most important interactions in nature, namely the hydrogen bond. The hydrogen bonding gives rise to extremely rich and complex phases in ice (Fig. 1)[7,8,9] even at not very high pressures and temperatures (that is < 2GPa and < 300K): 14 crystalline forms and several amorphous states[10,11] (low-density (LDA), high density (HDA), and very high density amorphous ice (VHDA)) have been experimentally determined. Hence this makes ice/water an excellent candidate for studies of common principles or regulations of phase transitions. Using pressure the oxygen-oxygen distance can be continuously modified, offering a possibility for understanding the hydrogen bond in a controlled way. On the basis of the behavior of water-containing systems, there was a concept about the symmetrization of the O-H…O bond, which implies that the hydrogen atoms would sit midway between the two neighboring oxygen atoms beyond the pressure when the O-O distance was less than about 2.4Å.[12,27] In addition, as to higher-pressure behavior there has been a number of intriguing suggestions, such as the metallization.[13] However, the behavior of ice at high pressure is still largely unknown. Beyond 62GPa, the only phase now experimentally discovered is ice X.[14,15,16] To the best of knowledge, the highest pressure explored to date on ice is 210GPa.[17] But a very broad P-T phase space whatever we interest with can be explored by using computer simulation. Based on *ab initio* molecular dynamics, M. Benoit et al.[1] proposed a higher-pressure phase ice XIII$^M$ (because of the two newly experimentally discovered phases of ice XIII and ice XIV in other P-T space,[18] we add a superscript "M" here for differentiation), which has a distorted hcp substructure of oxygen atoms.

In this paper, we discover a fine structure corrected for ice XIII$^M$, and conventionally nominate it as ice XV. Furtherly, we investigat the variations of hydrogen bonding, geometry and electronic structure among the ice X phase (55GPa~380GPa), ice XIII$^M$ (380GPa), ice XV (300–380GPa), as well as the ambient pressure phase ice XI,[19,20,21,22,23,24,25,26] which has a quasi-hcp substructure of oxygen, for comparison. Our results show that the hydrogen bonding structure and electronic structure of ice under high pressure are quite different from those under ambient pressure. Different from many other materials, the band gap of ice X increases linearly with pressure from 55GPa up to 290GPa. We explained this behavior based on Mulliken population analysis, the partial density of states (PDOS) and the charge transfers. The stable pressure range of ice X is confirmed to be from about 75GPa to 290-300GPa. Beyond 300GPa, we find a new structure featuring in broken symmetry in hydrogen bonding, namely the ice XV. . The geometry structure of ice XV shows that the hydrogen atoms may not just locate in the mid-point of two neighboring oxygen atoms under very high pressure, questioning the traditional symmetric hydrogen-bonded model[27,28,29,30,31,17] for ice under sufficient compression. The detailed investigation on the electronic structures and population analyses of both ice X and ice XV indicate that ice is a good wide gap insulator even up to 380GPa, and the metallization is not likely to happen in both phases. The stability of ice XV is extensively evaluated from both the enthalpy and lattice dynamics. The physical mechanism within the phase transformation from ice X to ice XV, instead of ice XIII$^M$, is revealed by the asymmetric repelling force acting on two hydrogen atoms within every tetrahedral coordination unit in the process of inter-sliding of oxygen layers under compression.

The organization of this paper is as follows: the *ab initio* method and the technique we use in our calculations are described in Sec. II. The results for ice under different high pressure and the detailed discussions are presented in Sec. III. Finally, conclusions are drawn in Sec. IV.

## II. METHOD

We perform the *ab initio* pseudopotential plane wave calculations based on the density functional theory (DFT) with norm-conserving pseudopotentials by the CASTEP code.[32] Exchange and correlation effects are treated by a generalized gradient approximation (GGA) with the Perdew-Burke-Eruzerhof (PBE) functionals.[33] The **k**-point grids for Brillouin zone sampling are generated via the Monkhorst-Pack scheme.[34] After convergence tests, we use 550eV cutoff of the kinetic energy and approximately 0.035Å$^{-1}$ separation of **k** points to generate Monkhorst-Pack mesh for different phases under different pressures. In addition to an ultrafine self-consistent field tolerance of $5.0 \times 10^{-7}$ eV/atom in the wavefunction optimization, and with energy convergence tolerance of $5.0 \times 10^{-6}$ eV/atom and geometry convergence tolerance of $5.0 \times 10^{-4}$ Å during the geometry optimizations, a finite basis set correction for the evaluation of energy and stress is applied. A linear interpolation scheme developed by Ackland[35] is used for evaluation of DOS. The atomic charges and bond populations are calculated by Mulliken population analysis (MPA) technique, which projects plane wave states onto a linear combination of atomic orbitals basis set.[36] The phonon dispersions and density of phonon states are calculated by PHONON[37] and CASTEP codes. These methods have been widely used.[38][39][40]

## III. RESULTS AND DISCUSSIONS

### A. The crystal structures of high-pressure phases of ice

The structural parameters of ice X, ice XI, ice XIIIM and ice XV at representative pressures are presented in Table I, and the 3D structures of their conventional unit cells are shown in Fig. 2[c].

### 1. The crystal structure of ice X

In ice X, all "$H_2O$ molecules" are dissociated to form an atomic crystal, Fig. 2b. This phase has been called 'proton-ordered' symmetric ice, since all protons are at the midpoints of O-O pairs. The lattice structure of ice X is primitive cubic with space group $Pn\bar{3}m$. It has a bcc substructure of oxygen atoms, and contains two $H_2O$ units in one unit cell.

---

[c] The 3D structures of ice XV and ice XIII$^M$ looks very similar, so we put them in one figure, i.e. Fig. 2c.

## 2. The crystal structure of ice XIII$^M$

Using *ab initio* molecular dynamics (Car-Parrinello Molecular Dynamics, the CPMD code[41 42]) for 16 H$_2$O molecules, M. Benoit et al. predicted that ice X would transform into a new phase ice XIII$^M$ (Ref.1) beyond 300GPa in 1996, but without any experimental confirmation yet. Ice XIII$^M$ is orthorhombic with four H$_2$O molecules per primitive cell, as is shown in Fig. 2c. The space group is *Pbcm* with a distorted hexagonal-close-packed (HCP) substructure of oxygen atoms, because the lengths of the six sides are not exactly equal.

## 3. The crystal structure of ice XV

Our calculation for ice at ground state (0 K) shows that there are more complicated hydrogen bonding structures in ice XIII$^M$ than the previous reported. A fine structure featuring in broken symmetry in hydrogen bonding in ice is discovered beyond 300 GPa. We nominate the new fine structure of ice XIII$^M$ as ice XV. Their 3D structures look very similar, and both are shown in Fig. 2c. With the same symmetry operators of space group *Pbcm*, both ice XV and ice XIII$^M$ have three inequivalent atomic sites: one oxygen site and two hydrogen sites. Because the *x* and *y* coordinates of oxygen are not in special position and vary under different pressure, while the value of which could not be exactly compared using different pseudopotentials and within different *ab initio* programs, we will only consider the difference between them in their hydrogen bonding structure in the following discussion. One apparent difference is the broken symmetry of one inequivalent hydrogen site in ice XV. The quantitative comparison of coordinates of their three inequivalent atomic sites under several representative pressures at 0 K is present in

Table I. The $x$ component of the fractional coordinate of that hydrogen in ice XIII[M] is always in a special position (0.0), while in ice XV it is ordinary and increasing with pressure, from 0.0020 at 300GPa to 0.0033 at 380GPa and without any evidence to stop[d]. More detailed discussion about the fine bonding structure in ice XV is present in Subsection III.C, and the underlying physical reason for this fine bonding structure is analyzed in Subsection III.E.

*4. The crystal structure of ice XI*

For comparison with the case of ice under ambient pressure, we also investigate the low-pressure low-temperature phase ice XI (Fig. 2a), the ordered form of ice $I_h$. Ice XI is also orthorhombic with space group *Cmc2₁*. Ice XI has a quasi-hcp substructure of oxygen, because the six sides are not exactly in one single plane despite they have equal lengths.

Comparing the lattice parameters of ice XI from our theoretically prediction with the experimental data in Table I, we confirm that our calculation could reproduce the real structure of ice well, especially we notice that the values of b/a and c/a in our model are closer to the ideal hcp structure than the two experimental data now available. However, it should be pointed out here that the GGA functional tends to underestimate the absolute lengths of lattice vectors or overestimate the pressure. This could be verified by the lengths of lattice constant *a* shown in both XI at 1atm and ice X at 62GPa in Table I.

---

[d] The convergence tolerance of maximum atomic displacement in our calculation is $5 \times 10^{-4}$ Å. Hence, the variation in $x$ coordinate of this inequivalent hydrogen should be counted.

**B. The stability of ice X and possible phase transformations under high pressure**

The pressure dependences of the lattice parameters, density, bond lengths (or the nearest neighboring atomic distances), enthalpy, Fermi energy, band gap, and the frequencies of phonon modes at Γ point are illustrated in Figs. 3-5, respectively.

*1. Ice X under pressure among 75GPa - 290GPa*

Between 75GPa and 290GPa in Figs. 3-4, we could see that all quantities vary smoothly and have simple linear relationships versus the pressure. The increase of Fermi energy is only due to the work done by external pressure, because we do not take account of any temperature effect. The band gap varies from 7.0eV at 75GPa to 10.5eV at 290GPa. This indicates that ice X is a good wide gap insulator, and the metallization is not likely to happen at current pressure range.

*2. Ice X under pressure below 75GPa*

However, examining the pressure dependence of neighboring atomic distances carefully in Fig. 3c, we could find that the changing slope below 75GPa is different with that above 75GPa: the smaller/larger gradient above/below 75GPa indicates ice is more difficult/easier to be compressed. In Fig. 5, we find that the frequency of acoustic phonon mode at Gamma point vanishes at 75GPa, and the entire acoustic branch goes soft below 75GPa. These evidences indicate that the lattice structure of ice X would be mechanically unstable below 75GPa, and the transition pressure to ice X is about 75GPa. This is in good agreement with the experimental result from the Raman spectra of ice,[43] which indicates the transformation occurs at about 70GPa for $H_2O$ and 85GPa for $D_2O$, and the result from the extrapolation of the infrared data of Aoki et al,[44] which predicted to be about 62GPa at ambient temperature.

### 3. Ice X under pressure above 290GPa

There is an obvious singularity at 290-300GPa, where the band gap stops the linear increasing with the pressure and suddenly getting to be flat. The possible phase transformation is also suggested by the *ab initio* molecular dynamics simulation carried out by M. Benoit et al.,[1] who proposed that a new ice XIII$^M$ phase would appear beyond 300GPa.

## C. The fine bonding structure and Mulliken population analyses of ice

### 1. The broken symmetry of hydrogen bonding in atomic crystal ice XV

The hydrogen and oxygen atomic distances and the bond populations in ice under different pressures are presented in Table III. From ice X to ice XIII$^M$/XV, the oxygen substructure transforms from body-centred-cubic (BCC) to distorted HCP structure, thus the oxygen-oxygen coordination number increases from 8 to 12. In ice XIII$^M$, the positions of the atoms were reported[1] to be O at ($u$, $v$, 0.25), H1 at (0, $v$-0.25, 0.25), and H2 at (0.5, 0.5, 0), where $u$ and $v$ are independent internal structural parameters. Based on the symmetry operators of *Pbcm*, the other symmetry equivalent position of oxygen with $z$ coordinate equals 0.25 is at (-$u$, $v$-0.5, 0.25). Thus H1 always locates in the middle of two neighboring oxygen atoms, and so does H2. Hence, as the same to ice X, the 'proton-ordered' symmetric ice phase, all hydrogen atoms in ice XIII$^M$ also locate in the symmetric positions of the corresponding O-H-O bonds, and there should be only two different neighboring H-O bond lengths, despite the regular-tetrahedron coordination structure in ice X is now distorted, as is shown in Fig. 6a.

However, examining the bond lengths of ice XV in Table III, we can find that there are three different lowest values for the four H-O bond lengths in the fourfold coordination structure, e.g. 0.9977Å, 1.0105Å, and 1.0213Å at 380GPa[e]. This is because there are two bonds having equal lengths, describing two hydrogen atoms locating in the middle of two neighboring oxygen atoms; The other two bonds having shorter or longer lengths describe the other two hydrogen atoms which have asymmetric positions between the neighboring oxygen atoms, which are shown in Fig. 6b.

For making sure this bonding feature in the tetrahedral coordination structure of ice under high pressure, we have further optimized several other geometries of ice under 350GPa, including ice XI, distorted structures of ice VII and ice VIII, and find that the feature of broken symmetry in hydrogen bonding do commonly exist. Moreover, we check our calculations by re-optimizing the ice XIII$^M$ structure using other exchange-correlation functionals: all of the three types of GGA functionals (PBE[33], RPBE[45], and PW91[46]) and CA-PZ[47] type of LDA functional implemented in the CASTEP code and the PAW[48] pseudopotential. All of these calculations have reached the same result: 50% symmetric hydrogen-bonded structure and the other 50% asymmetric. This fine bonding feature in ice XV is different from ice XIII$^M$, the previous result reported by M. Benoit et al. (Ref.1), despite they have the same space group. Hence, our results suggest that further researches, especially experimental works, are needed to recheck the long-standing hydrogen bonding model in ice under very high pressure considering that all hydrogen atoms should be equilibrated in the midway of two neighboring oxygen atoms beyond the pressure when the oxygen-oxygen distance $d_{OO}$ is less than about *2.4* Å (Refs. 12, 27, 28, 29, 30, 31, 17), as happens in atomic crystal ice X (Fig. 6c).

---

[e] The convergence tolerance of maximum atomic displacement in our calculation is $5 \times 10^{-4}$ Å. Hence, the length difference in the asymmetric O-H-O bond, $\sim 10^{-2}$Å, is much bigger than our calculation inaccuracy.

*2. Abnormal atomic distance variation in ice XV under compression*

For the bond lengths (or neighboring atomic distances), it is normal that most bonds in both ice X and ice XV shrink under stronger compression, nevertheless, we find that there is an exception in one distance of H-O pair in ice XV. Under higher pressure, the longest neighboring H-O distance that listed in Table III for ice XV extends instead, from 2.1907Å under 300GPa to 2.2333Å under 380GPa. In fact, this is due to the tendency to form an ideal *ABAB* stacking of HCP substructure of oxygen lattice, leading the relative sliding of neighboring oxygen layers perpendicular to [001] direction along the [010] direction or reversely, and the exceptional oxygen and hydrogen atoms above belong to these two layers, respectively. The evidence from the displacements of the corresponding hydrogen atom and oxygen atom are presented in Table IV.

*3. The variation of charge transfers and bond populations of ice under high pressure*

We present for the three phases the effective charge densities of H 1*s*, O 2*s*, and O 2*p* states, and the charge transfers of hydrogen atoms (corresponding to half of those of oxygen atoms) in Table II, as well as the neighboring atomic distances and bond populations in Table III. Though as we all know that the absolute values of the atomic charges yield by the population analysis have little physical meanings because of their high sensitivity to the atomic basis set, we are still able to find useful information from the relative values of Mulliken populations. The bond populations indicate the overlap degree of the electron clouds of two bonding atoms, and the lowest/highest values imply that the chemical bond exhibits strong ionicity/covalency, respectively. The negative values of bond populations for H-H and O-O bonds indicate that these bonds have the tendency to be broken or these atoms have Van der Waals interaction only.

Analyzing the charge transfers from hydrogen atoms to oxygen atoms and bond populations for ice X and ice XV under different pressures listed in Table II and III, respectively, we find that in both phases the charge transfers are decreased while the absolute values of bond populations are increased with the increase of the pressure. This could be understood that more electron clouds overlap together and the attractive effect between the electrons and protons become obvious, thus less charge could be transferred from hydrogen to oxygen when the H-O bond length is compressed to shorter. In addition, due to the increase of the bond population, the covalency character in both ice X and ice XV strengthens under higher pressure. Hence the metallization is not likely to happen in both two phases. It is also interesting to notice that under 300GPa the neighboring H-O bond populations of ice XV (~0.49) are quite different with ice X (~0.37) but similar to ice XI (~0.49), implying that the ionicity of ice XV under high pressure is very different with ice X but close to ice XI under ambient pressure. On the other hand, considering the change of bond population induced by the pressure within structures having same symmetry, e.g. in ice X from 75GPa compressed to 300GPa, we find that the H-O bond population increases only a little, from 0.35 to 0.37. Meanwhile, we remind that ice XV and ice XI have similar packing model close to HCP. Hence, we infer that the symmetry of the structure of ice would be a much more significant factor affecting its ionicity.

**D. Electronic structures of ice under high pressure**

*1. The charge density distribution of ice – from molecular crystal to atomic crystal*

Fig. 7 shows the spatial charge density distributions of ice XI, ice X, and ice XV in a (100) plane, a (110) plane, and a (001) plane, respectively, all of which pass through at least one $H_2O$ molecular plane within their unit cells. In Fig. 7a, we can clearly identify the H-O…H bonds and $H_2O$ molecules in ice XI. However, in either Fig. 7b or Fig. 7c for ice X and ice XV, it is difficult to find any separated molecules, because all $H_2O$ molecules have been dissociated to form the atomic crystals. In addition, we can find many similarities between Fig. 7b and Fig. 7c. This is due to the fact about the formation of ice XV from ice X as we analyzed in the subsection above: one of the two neighboring layers (that is $(110)_X$ plane, i.e. $(001)_{XV}$ plane in ice XV) in ice X sliding along the $[001]_X$ (i.e. $[010]_{XV}$) direction.

*2. The band structures of ice XI, ice X, and ice XV*

Because the band structure of ice X changes slightly in the pressure range we investigate, we will only present the result under 300GPa here for illustration. From Fig. 8, we can find all of the three phases are insulators with direct band gap at highly symmetric Γ point. It is interesting that ice XV has a little smaller band gap (~10.36eV) than that of ice X (~10.50eV) under 300GPa, and both are larger than that of ice XI (~5.15eV) under ambient pressure (here the band gap ~5.15eV is a little smaller than the experimental value ~7.8eV for ice $I_h$ at ambient condition because of the well-known reason that the GGA tends to underestimate the band gap). In Fig. 8a, we can see that the valence bands (VBs) of ice X consist of 8 levels and degenerate to 4 at all high symmetric **k**-points; while in Figs. 8b and 8c, ice XV and ice XI have 16 levels in the VBs, because they have double number of $H_2O$ molecules within the unit cells than ice X. The lattice structure of ice XV (8 symmetry operations) has lower symmetric operators than ice X (48 symmetry operations), the valence bands only degenerate to four levels at **k**-points U and R. The valence bands of ice XI are discrete in energy as most molecular crystals, where most valence bands are flat and seldom overlapped. They degenerate from sixteen levels to six levels at **k**-points Z, T, and R, because there are six molecular electronic energy levels of $H_2O$.

### 3. The partial density of states and the hybridization of atomic orbitals in ice under high pressure

From our detailed *ab initio* calculations for ice X under a large pressure range (55-380GPa), we find that the DOS spectra have a red shift below Fermi energy and a blue shift above the Fermi energy when the pressure is increased. In Fig.9, we present the partial density of states (PDOS) for ice X under 75GPa and 300GPa, together with ice XV under 300GPa and ice XI under ambient pressure for illustration. To trace the origin of the complex DOS spectra for the two high pressure phases, ice X and ice XV, both of which are atomic crystals, we start from the analysis of the DOS spectra of ice XI, which phase is molecular crystal, as illustrated in Fig. 9a. There are six peaks below Fermi energy, corresponding to six molecular electronic energy levels of $H_2O$ from the lowest to the highest: $\sigma_s, \sigma_s^*, \sigma_x^{non}, \sigma_z, \sigma_z^*$ and $\pi_y^{non}$ (here marks with no superscript, with superscript "*" or "non" represent the bonding, anti-bonding or non-bonding state, respectively). Within the VBs, O 2$s$ and H 1$s$ contribute to the formation of $\sigma_s$ and $\sigma_s^*$, O 2$p_x$ and H 1$s$ to $\sigma_x^{non}$, O 2$p_z$ and H 1$s$ to $\sigma_z$ and $\sigma_z^*$, and a lone-pair O 2$p_y$ orbital to $\pi_y^{non}$, respectively. The conduction bands (CBs) in 14.6-16.2eV are contributed by O 2$p$ and H 1$s$ orbitals, while the lowest unoccupied molecular orbital, between ~5.15eV and ~14.6eV, is mainly from the $s$ states of hydrogen. Turning to Figs. 9b and 9c for ice X, we can find that the DOS spectra become complex due to increase in band dispersion. For ice X under 300GPa, the CBs in 16.5-23.5eV originate mainly from the O 2$p$ and H 1$s$ states, and those between 14.9-16.5eV are from the cooperative contributions by O 2$p$, O 2$s$, and H 1$s$ states. The CBs below ~14.9eV are mainly contributed by an anti-bonding orbital from $s$ states of oxygen and hydrogen. As in the case of ice XI, the two peaks between ~-30.0eV and ~-15.5eV having the lowest energies in the DOS spectra of ice X are mainly the bonding and anti-bonding states from the $s$ states of oxygen and hydrogen. The VBs between ~-15.5eV and ~-5.7eV come from the cooperative contributions of O 2$p$, O 2$s$, and H 1$s$ states. The electronic states above ~-5.7eV and below the Fermi energy are dominantly from O 2$p$ states as in the case of ice XI. Based on the band gap among the valence bands (V-Bandgap), we divide the valence bands to two regions: VB1 and VB2, as shown in Fig. 9.

Comparing the DOS spectra of ice X under 300GPa and that under 75GPa, we can find that all peaks become broader and their heights drop with the increase of pressure, the peaks below the Fermi energy shift toward lower energies (red shift) while those peaks above the Fermi energy shift toward higher energies (blue shift). This feature accounts for the broader energy region for optical response, which we will discuss in another paper. In addition, we notice that the band gap around -15.5eV tends to disappear under higher pressure, indicating more bands overlapping and remarkable band dispersions in this energy region. Furthermore, it is interesting that the band gap is increased by higher pressure, from 7.04eV at 75GPa to 10.50eV at 300GPa, thus the metallization should be impossible in this ice phase. This anomalous behavior of pressure dependence is due to the electronegativity of oxygen in ice X is relatively decreased (the evidence is the charge transfer from oxygen to hydrogen) and the H-O bond population is increased, resulting in weaker ionicity (stronger covalency) and larger band gap, unlike many other materials under high pressure. The comparison of the DOS spectra of ice XV with ice X indicates that they have quite similar profile of the electron energy dispersion relations in total *s* states, total *p* states, and their summation, except the partial DOS spectra of *s* states from specific atoms and the more peaks appear in ice XV.

The integration of the DOS spectra is useful for understanding the hybridizations of atomic orbitals and the charge transfers calculated from Mulliken population analysis in Table II. Comparing the values at 0eV in Figs. 10a-c, we find that in the three phases of ice the ratios of total *s*/*p* states are very close to 3/5, as the $sp^3$ hybridization in $H_2O$

molecule. Those are 11.475/19.321, 5.936/9.961, and 11.900/19.906, in ice XI, ice X, and ice XV, respectively. However, it is interesting to notice that the *s/p* ratio is increased with the pressure, from 0.594 in ice XI, 0.596 in ice X, to 0.598 in ice XV. This variety is arising from the charge transfers of *s* electrons both from hydrogen and oxygen to O 2*p* states. For instance, the integration from the lowest energy to 0eV for the densities of *s* states respectively from hydrogen and oxygen are 4.29 and 7.18 in ice XI, 2.70 and 3.23 in ice X, and 5.42 and 6.49 in ice XV, so they make the difference of *s/p* ratio. On the other hand, the charge transfer per hydrogen atom in ice X could be calculated as $(4 - 2.70)/4 \approx 0.33$. The value for oxygen is equal to the double that of hydrogen, i.e. 0.66. They are in good agreement with the values listed in Table II. The charge transfers for the other two phases could also be calculated in the same way. In fact, the definite integration within certain energy region of the DOS spectra could be used to analyze the hybridization of atomic orbitals. For instance, the ratios of the number of total *s* states versus total *p* states in two regions (~-6.5eV to ~-5.4eV) and (~-4.2eV to ~-1.2eV) in ice XI are both close to 1/3, thus the VBs in these two energy regions are very likely dominated by *sp$^3$* hybridization.

**E. The stability of ice XV**

*1. Could ice XV be more stable?*

We firstly judge through the ground-state enthalpies of ice X, ice XV, and ice XIII$^M$ at 0K among 280-420GPa$^f$, as shown in Fig. 11. Apparently Ice XV is energetically the most favourable structure beyond 290GPa. Although the energy difference between ice XV and ice XIII$^M$ is small at about 300GPa, it comes to be bigger and bigger under higher pressure.

In order to check the mechanical stability of the structure of ice XV, we have done lattice dynamics for ice XV under different pressures. The phonon dispersions and the density of phonon states of ice XV under 300GPa and 380GPa are presented in Figs. 12 and 13, respectively. Because there are 12 atoms in the primitive cell of ice XV, there must be totally 36 phonon modes (the degenerate modes are also counted). From 300GPa to 380GPa, all phonon modes stiffen some with pressure, and there is no evidence revealing the tendency of softening of any phonon modes. Hence, we could safely conclude that ice XV should be more stable beyond 290GPa to at least 380GPa.

*2. Why ice X tends to transform to ice XV beyond 290GPa?*

The detailed population analysis and the variation of the dispersion of partial density of states (PDOS) of ice under different pressures may indicate the reason for the abnormal pressure dependent behavior of band gap in ice X and why it tends to transform to a new phase beyond 300GPa. From ice XI at low pressure to ice X at high pressure in Fig. 9, we notice that the V-Bandgap becomes narrower and narrower with pressure, and finally closes at about ~300GPa. Namely the energy eigenvalues of H 1$s$ and O 2$s$ in VB1 turn to be closer to those of electronic states in VB2, thus making the charge transfer between them possible. As shown in Table II, when the pressure is increased, more and more O 2$s$ electrons are transferred to O 2$p$ in VB2, from 0.29 at 75GPa to 0.38 at 300GPa, and the charge transfer from H 1$s$ to O 2$p$ varies only little, hence more and more O 2$p$ characters appear. This strengthens the $sp^3$ hybridization in VB2 in ice under high pressure. The stronger $sp^3$ hybridization in the tetrahedral coordination structure makes the electronic transition between the valence bands and the conduction bands more difficult, i.e. the band gap widens. Nevertheless, when the pressure approaching to about ~300GPa, the incomplete-filled $sp^3$ hybridization bonding structure comes to be not strong enough to keep the regular tetrahedron coordination structure under higher compression. One of the natural result is the lattice structure, especially the oxygen substructure, would be distorted and packed to be denser. That is why the *ABAB* stacking structure of distorted HCP structure forms in ice XV/XIII$^M$. The comparison about the densities of them is presented in Table II.

---

$^f$ Since there is no specific coordinate information about ice XIII$^M$ below 380GPa in literature, and the positions of oxygen varies under different pressure, here we relocate all hydrogen atoms in ice XV to be symmetrical as ice XIII$^M$ for comparison in this subsection.

### 3. Why ice XV is more stable than ice XIII$^M$?

As we discussed in the subsubsection above and the subsubsection of III.C.2, under high pressure beyond 290GPa, there are inter-sliding along the [010] axis in ice XIII$^M$/XV between the neighbouring oxygen layers (they are (001) planes), leading to the *ABAB* stacking of the HCP phase. Simultaneously, the hydrogen atoms bridging those layers (H$_5$,

$H_6$, $H_7$, and $H_8$ in Fig. 14) should follow and also move along the $[010]$ or $[0\bar{1}0]$ direction. Now considering a reference frame keeping those bridging hydrogen atoms fixed, as shown in Fig. 14, the front oxygen layer containing $O_1$-$O_3$-$O_3$'-$O_1$' slide along $[010]$ direction, and the back oxygen layer (containing $O_4$-$O_2$-$O_4$'-$O_2$') relatively slide along $[0\bar{1}0]$ direction. This is accompanied by the movement of the hydrogen atoms within those layers (e.g. $H_1$, $H_3$, and $H_2$, $H_4$, respectively), however, due to the net repelling force from the bridging hydrogen atoms, they could not rigidly follow the oxygen atoms. This movement difference will become more significant when the repelling force is larger under higher compression. Take $H_3$' in Fig. 14 for example, $H_3$' is apparently nearer to ($H_5$, $H_6$, $H_5$', and $H_6$') than to ($H_7$, $H_8$, $H_7$' and $H_8$'), thus there is net repelling force acting on $H_3$' from them along the $[0\bar{1}0]$ direction, leading $H_3$' to be closer to $O_3$' than to $O_1$. This is the physical origin of the broken symmetry happening in the hydrogen bonding in ice XV. Hence, we consider the description on page 2935, Phys. Rev. Lett. **76** (1996) [Ref.1] which states "The hydrogen atoms follow rigidly the displacements of the oxygen atoms still sitting midway between the two neighboring oxygen atoms" is not accurate for the new phase of ice beyond 300GPa.

## IV. CONCLUSION

We have carried out detailed investigations on the hydrogen-bonding, geometry, electronic structures for the two high-pressure phases of ice (ice X and ice XV), and compared them with the ambient pressure phase ice XI using the *ab initio* pseudopotential density functional method. One of the most remarkable differences among them is that ice X, ice XIII$^M$, and ice XV are atomic crystals but ice XI is molecular crystal. This feature is clearly identified from their spatial charge density distributions. Especially, we evaluate the structure stability of ice X under a large high pressure range (55GPa-380GPa), and reckon two critical phase transition pressure to be about 75GPa and 290-300GPa. Beyond 300Gpa, we discover a fine structure corrected for the previous theoretically reported phase ice XIII$^M$ and name it as ice XV. Ice XV has a distinct bonding structure from the conventional hydrogen-bonding model for ice under high pressure. We find that the new phase ice XV has 50% protons in symmetric positions within O-H-O bonds and the other half being asymmetric under super high pressure. The physical reasons for that ice X is stable under the pressure larger than 75GPa is evaluated from the pressure dependence of the frequencies of phonon modes at Gamma point, and that the pressure-induced phase transition at about 300GPa is analyzed from the change of *s-p* charge transfer from hydrogen and oxygen. The studies on the density of electron states and band structures show that the band gap of ice X is increased with high pressure among 75-290GPa, and we explain this behavior from the delocalization of electrons in the valence bands and the change of *s-p* charge transfers between hydrogen and oxygen under high pressure. The metallization of ice is not likely to happen in both ice X and ice XV phases. The stability of ice XV is specially discussed in a separate subsection from the phonon dispersion data and the density of phonon states, as well as the comparison of the favor of enthalpy among other ice phases. The physical mechanism for the phase transformation from ice X to ice XV (instead of ice XIII$^M$) is revealed that following the inter-sliding of (001) planes of oxygen layers along either the $[010]$ or $[0\bar{1}0]$ direction, the unidentical displacements of hydrogen atoms within every tetrahedral coordination unit result in asymmetric repelling forces acting on therein two hydrogen atoms within the oxygen layers, and thus the broken symmetry in hydrogen bonding occurs, namely the ice XV forms.

## ACKNOWLEDGEMENTS

The authors are grateful to Prof. Warren Pickett (University of California, Davis) for many useful discussions. This work was supported by the National Natural Science Foundation of China under grant No. 10574053 and 10674053, 2004 NCET and 2003 EYTP of MOE of China, the National Basic Research Program of China, Grant No. 2005CB724400 and 2001CB711201, and the Cultivation Fund of the Key Scientific and Technical Innovation Project, No. 2004-295.

# List of Figures

FIG. 1. Phase diagram of ice including some recently discovered stable and metastable forms (Lobban et al. 1998 (Ref. 49); Salzmann et al 2006 (Ref. 18)). Arrows indicate the typical preparation paths for low density amorphous ice (LDA), high density amorphous ice (HDA), ice XIII and XIV. (Ref. 50)

FIG. 2. (Color online) Crystal structures of (a) ice XI, (b) X, and (c) $XIII^M$/XV. Red (dark) and white circles denote oxygen and hydrogen atoms, respectively.

FIG. 3. (Color online) The pressure dependences of (a) the length of lattice constant *a*, (b) density, (c) bond lengths (or neighboring atomic distances) of ice X.

FIG. 4. (Color online) The pressure dependences of (a) enthalpy, (b) Fermi energy, and (c) band gap of ice X.

FIG. 5. (Color online) The pressure dependence of the frequencies of phonon modes at Gamma point.

FIG. 6. (Color online) The hydrogen bonding structures of (a) ice XV, (b) ice $XIII^M$, and (c) ice X. The atomic distance is in unit Å. The convergence tolerance of maximum atomic displacement in our calculation is $5 \times 10^{-4}$ Å.

FIG. 7. (Color online) Charge density slices of (a) ice XI (1atm) plotted in a (100) plane, (b) ice X (300GPa) plotted in a (110) plane, and (c) ice XV (300GPa) plotted in a (001) plane, all of which pass through at least one $H_2O$ molecular plane. The unit cell is outlined in blank. Corresponding to the arrangement of rainbow spectroscopy, from the zone of blue to red, the charge density is increased from the lowest to the highest.

FIG. 8. (Color online) Band structures of: (a) ice X under 300GPa, (b) ice XV under 300GPa, and (c) ice XI under ambient pressure.

FIG. 9. (Color online) Calculated spectra of partial density of states for (a) ice XI under ambient pressure, (b) ice X under 75 GPa, (c) ice X under 300GPa, and (d) ice XV under 300GPa.

FIG.10. (Color online) Integrations for the DOS spectra of (a) ice X under 300GPa, (b) ice XV under 300GPa, and (c) ice XI under ambient pressure.

FIG.11. (Color online) Enthalpies per H2O unit of ice X, ice XV, and ice XIII$^M$ at ground state (0K) are plotted as functions of pressure. The enthalpies are referenced to that of ice X. The energy convergence tolerance in our geometry optimizations is 5.0 x 10$^{-3}$ meV/atom.

FIG. 12. Phonon dispersions of ice XV (a) at 300GPa and (b) at 380GPa. The Brillouin zone of the orthorhombic lattice is shown at the right.

FIG. 13. Density of phonon states of ice XV (a) at 300GPa and (b) at 380GPa.

FIG. 14. 2x1x1 supercell structure of Ice XV at 380GPa. The measured bond lengths are in unit Å. The oxygen layer composed of ($O_1$-$O_3$-$O'_3$-$O'_1$) slides along the [010] direction, and the other layer composed of (O4-O2-$O'_2$-$O'_4$) oppositely slides along the [$0\bar{1}0$] direction.

**List of Tables**

TABLE I. Lattice parameters (a, b/a, c/a) and fractional coordinates of inequivalent atoms of ice XI, ice X, ice XV, and ice XIII$^M$ at ground state (0K).

TABLE II. Densities, charge transfers ($Q_{tran}$) of hydrogen atoms (corresponding to half of those of oxygen atoms), and effective charge densities ($Q_{eff}$) of H 1$s$, O 2$s$, and O 2$p$ states of ice XI, ice X, ice XV, and ice XIII$^M$ at ground state (0K).

TABLE III. Different neighboring atomic distances and bond populations in ice XI, ice X, and ice XV.

TABLE IV. The displacements of H3 and O4 which correspond to the elongated H-O bond in Table II for ice XV under stronger compression.

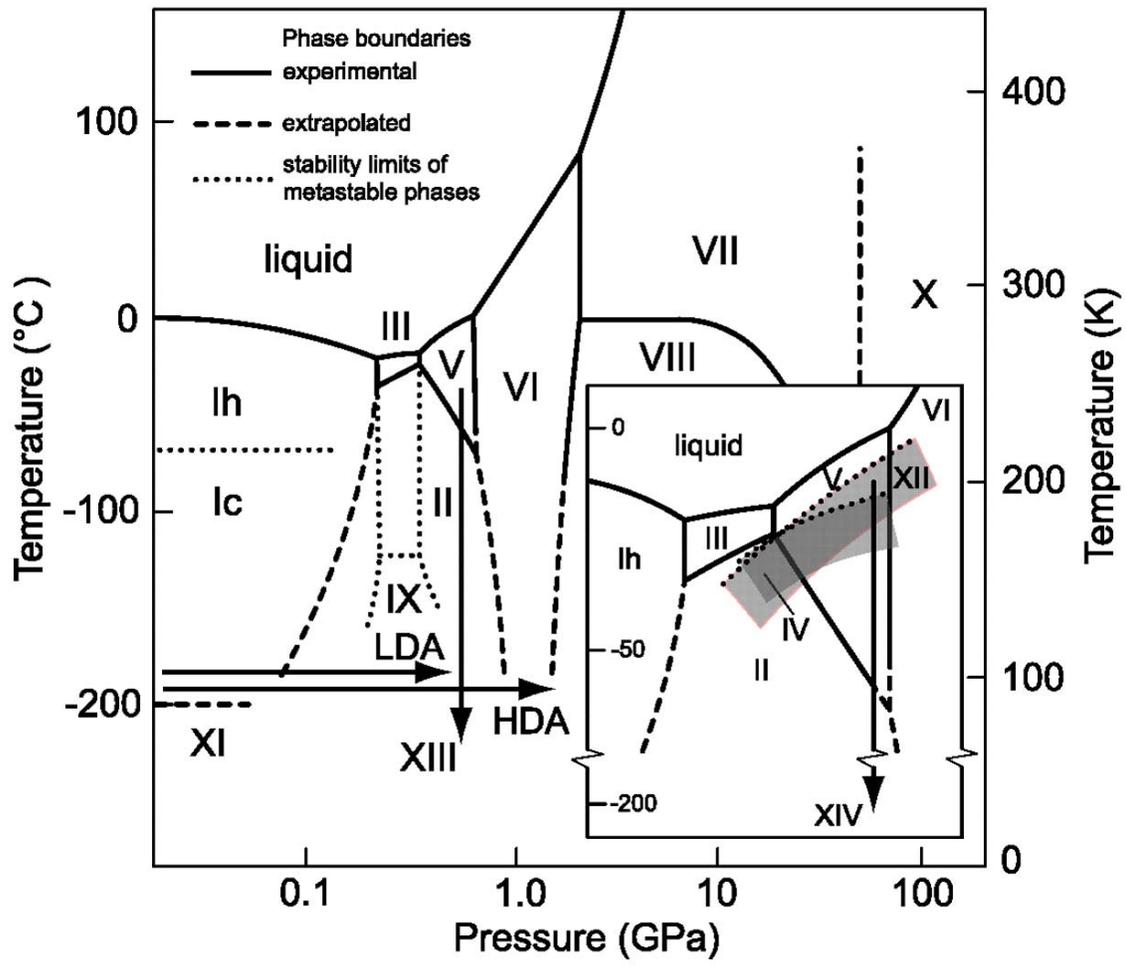

**FIG. 1**

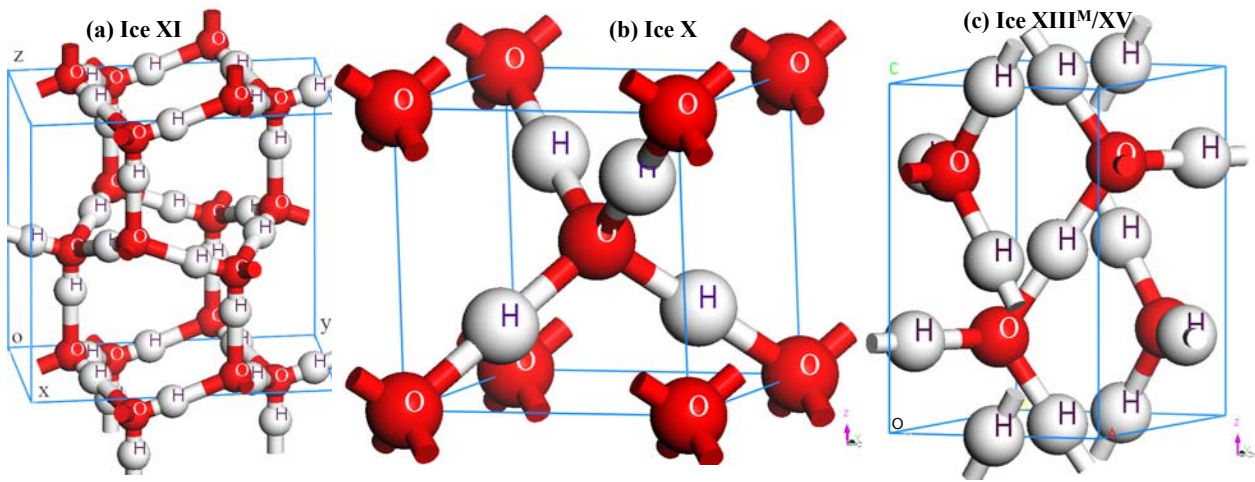

**FIG. 2**

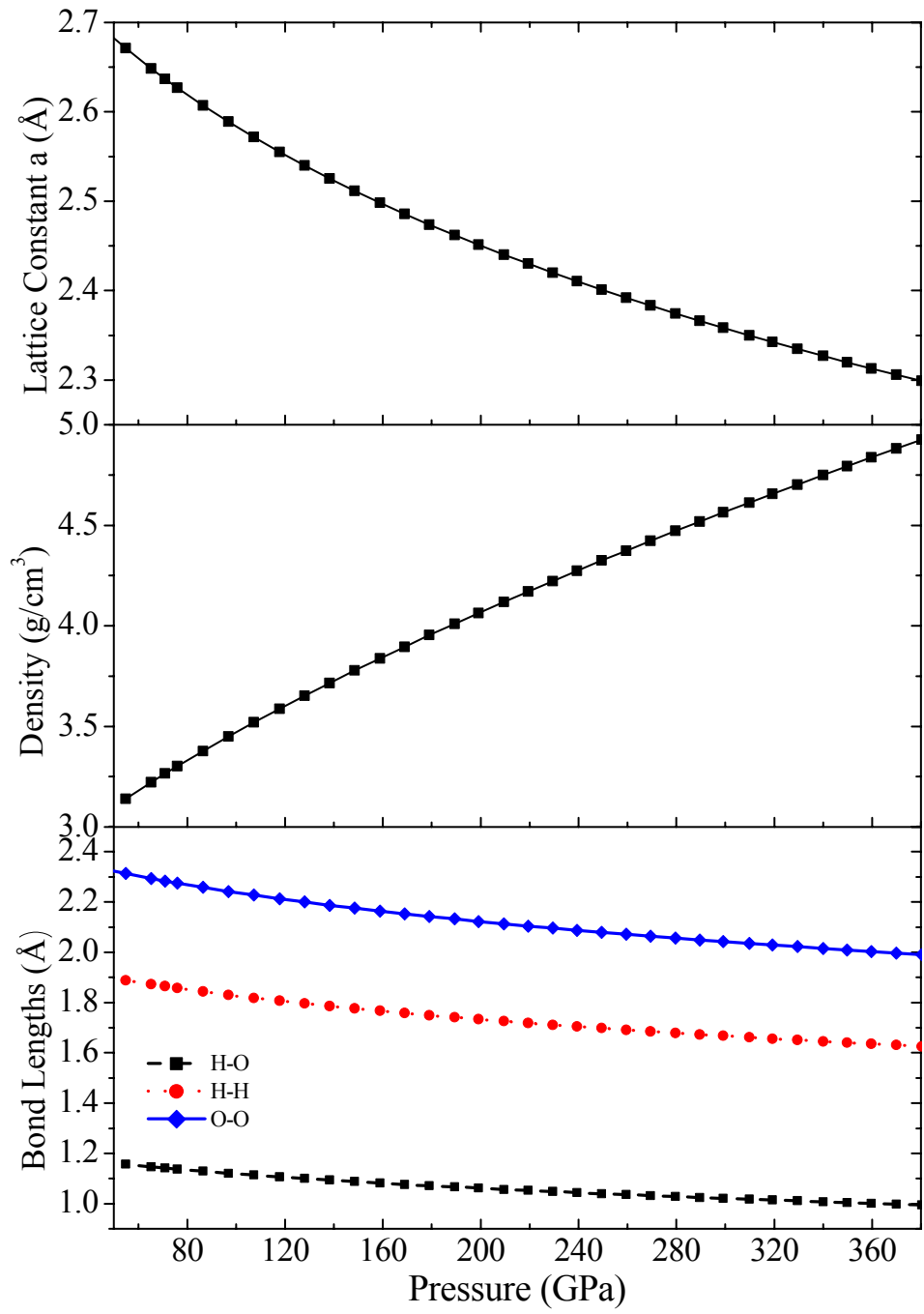

**FIG. 3**

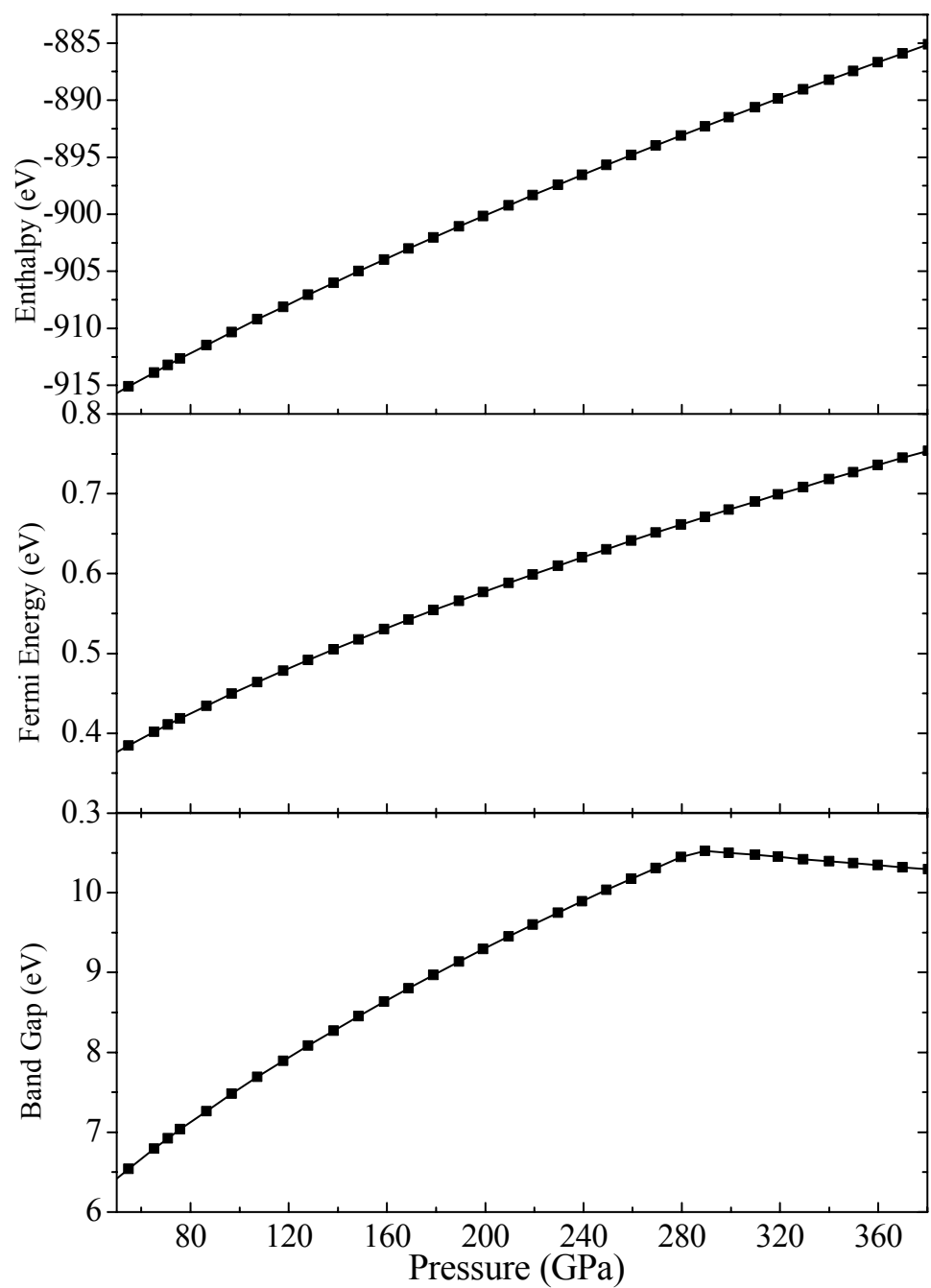

**FIG. 4**

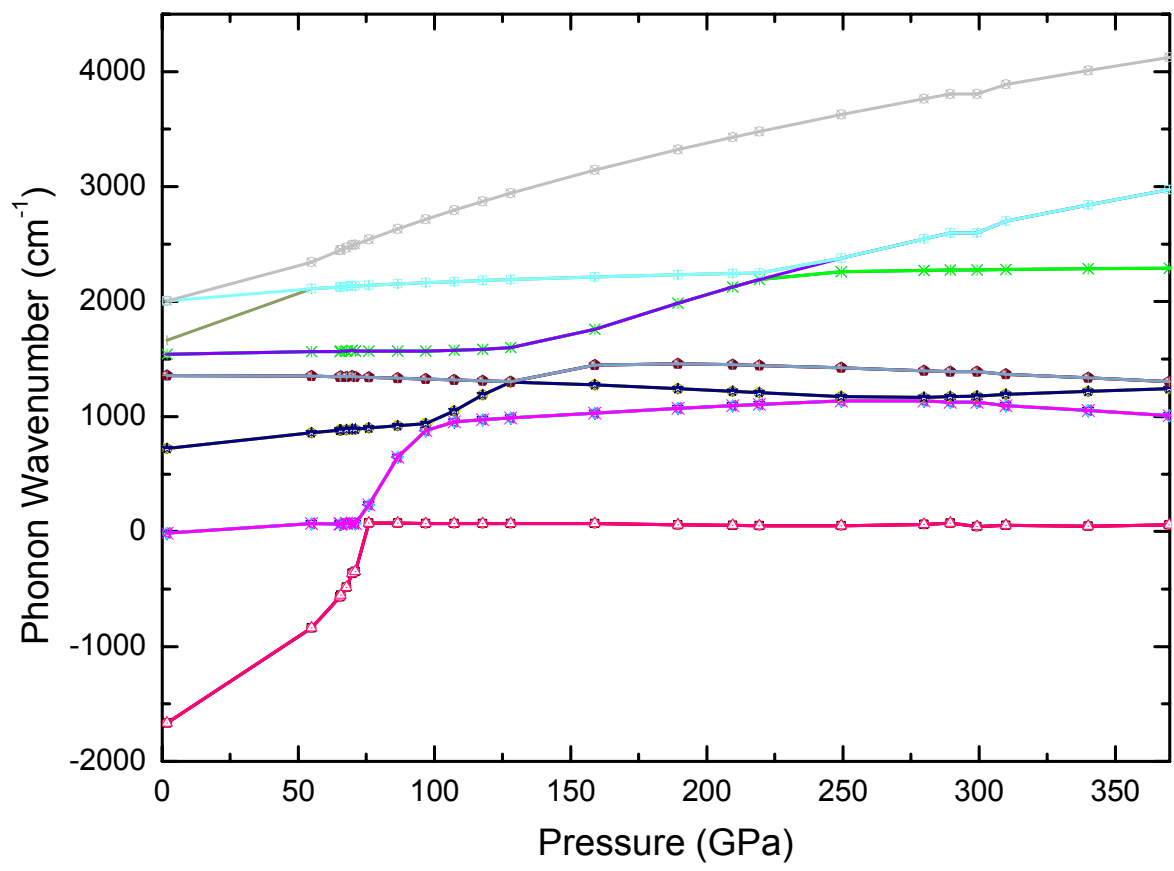

**FIG. 5**

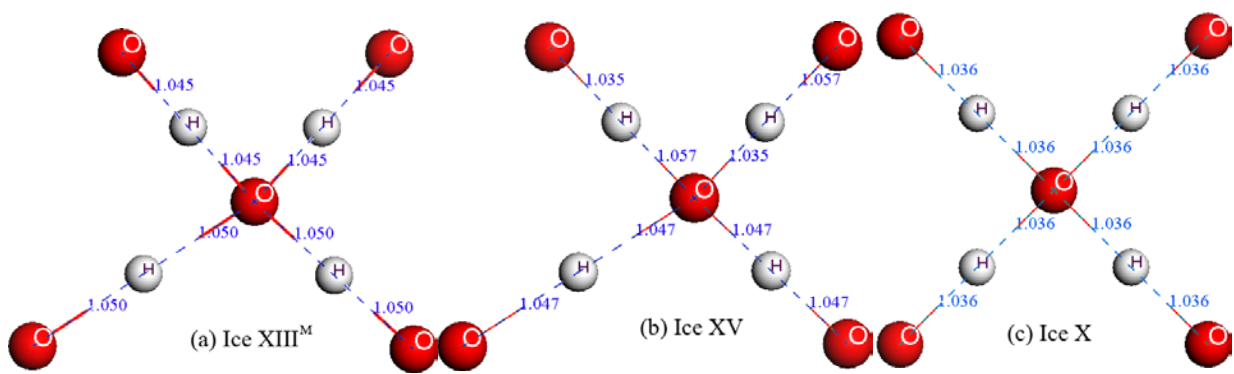

**FIG. 6**

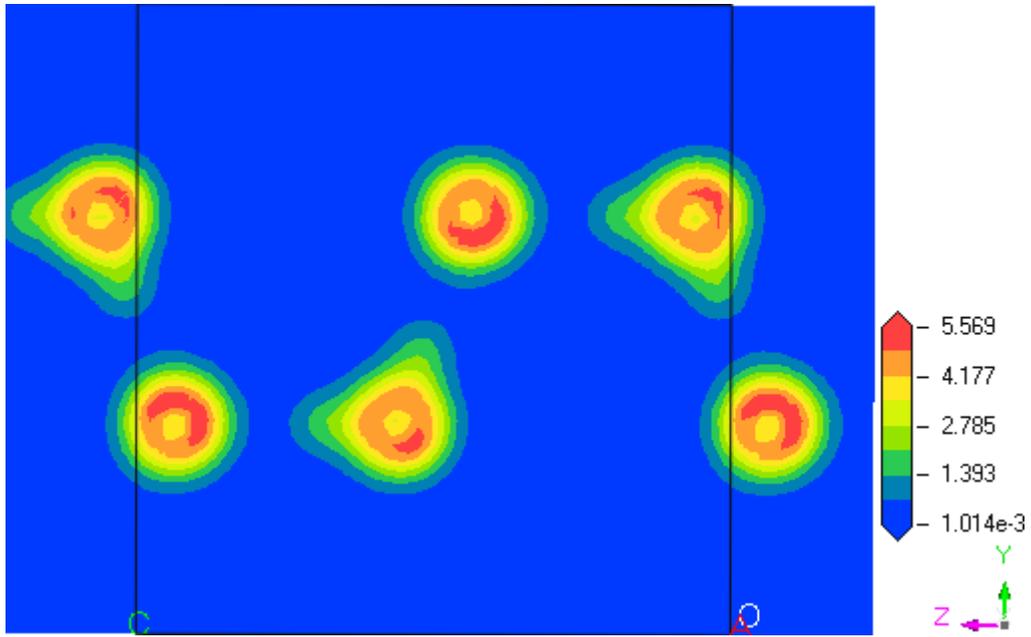

**FIG. 7a**

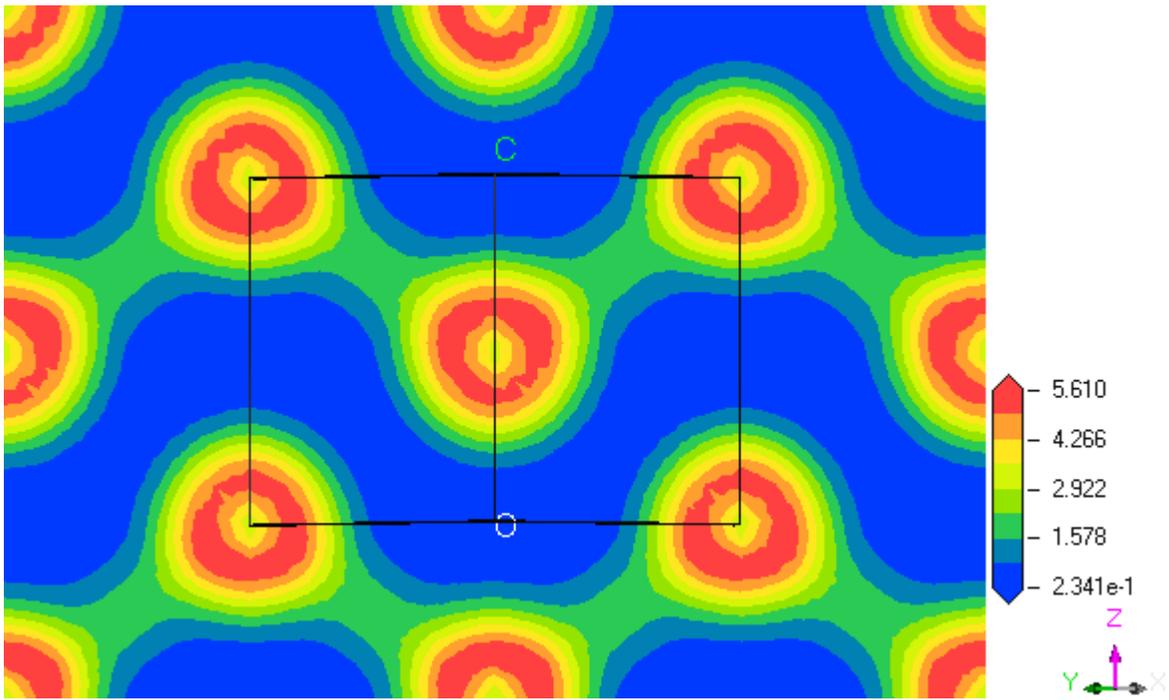

**FIG. 7b**

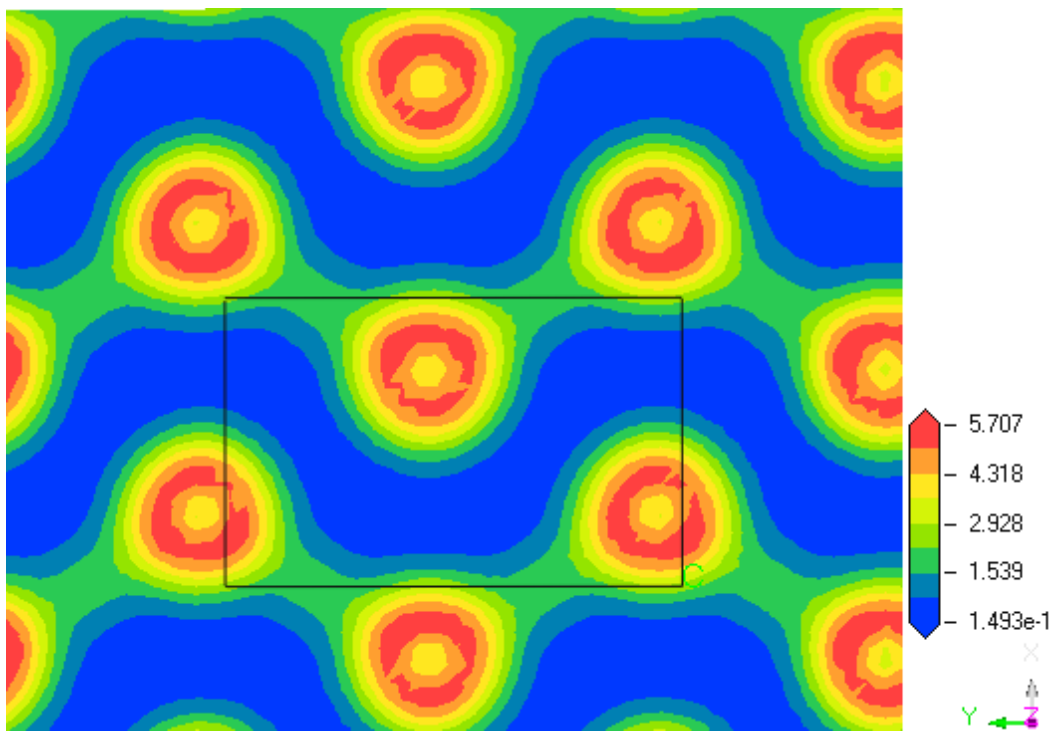

**FIG. 7c**

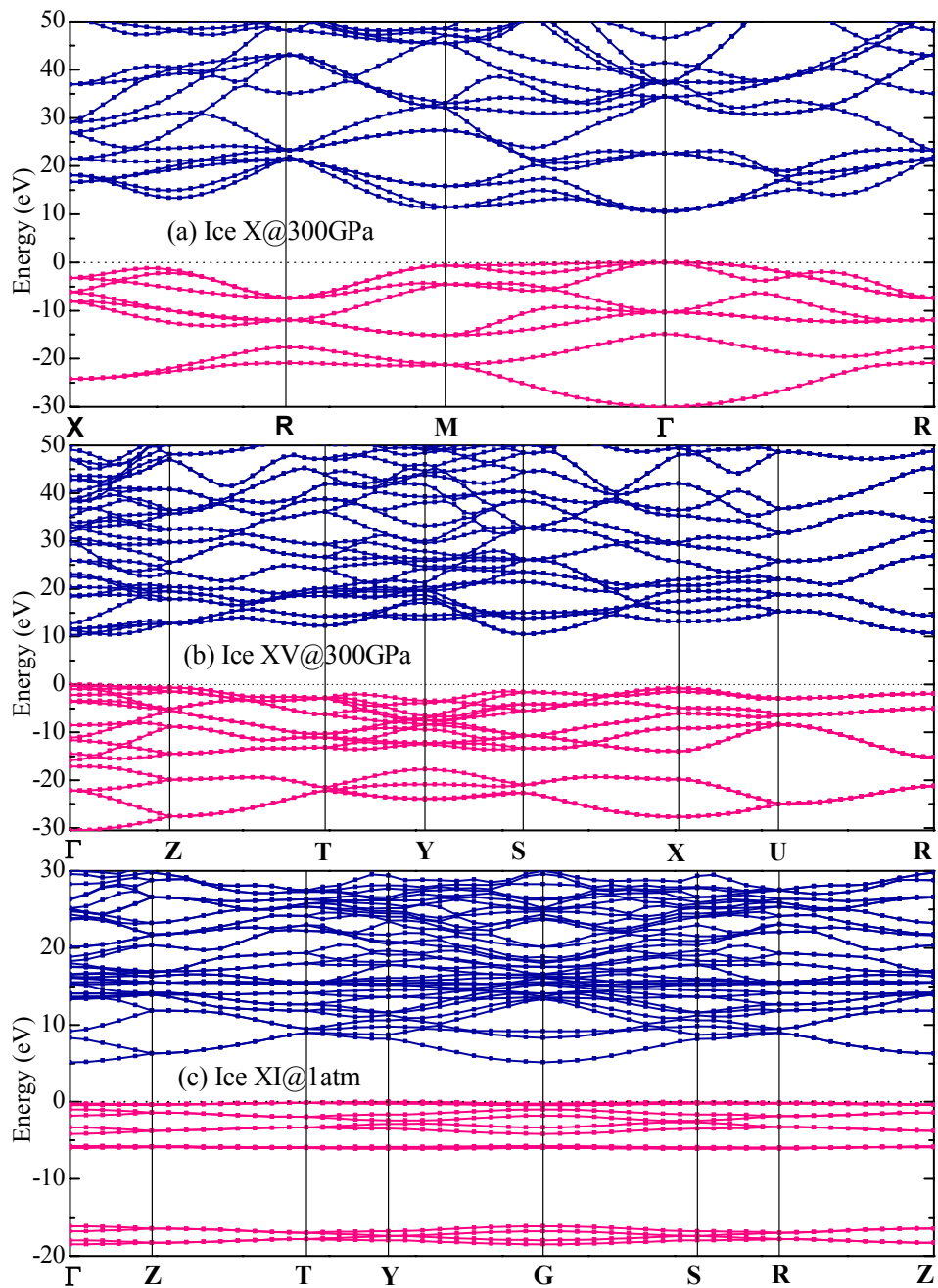

FIG. 8

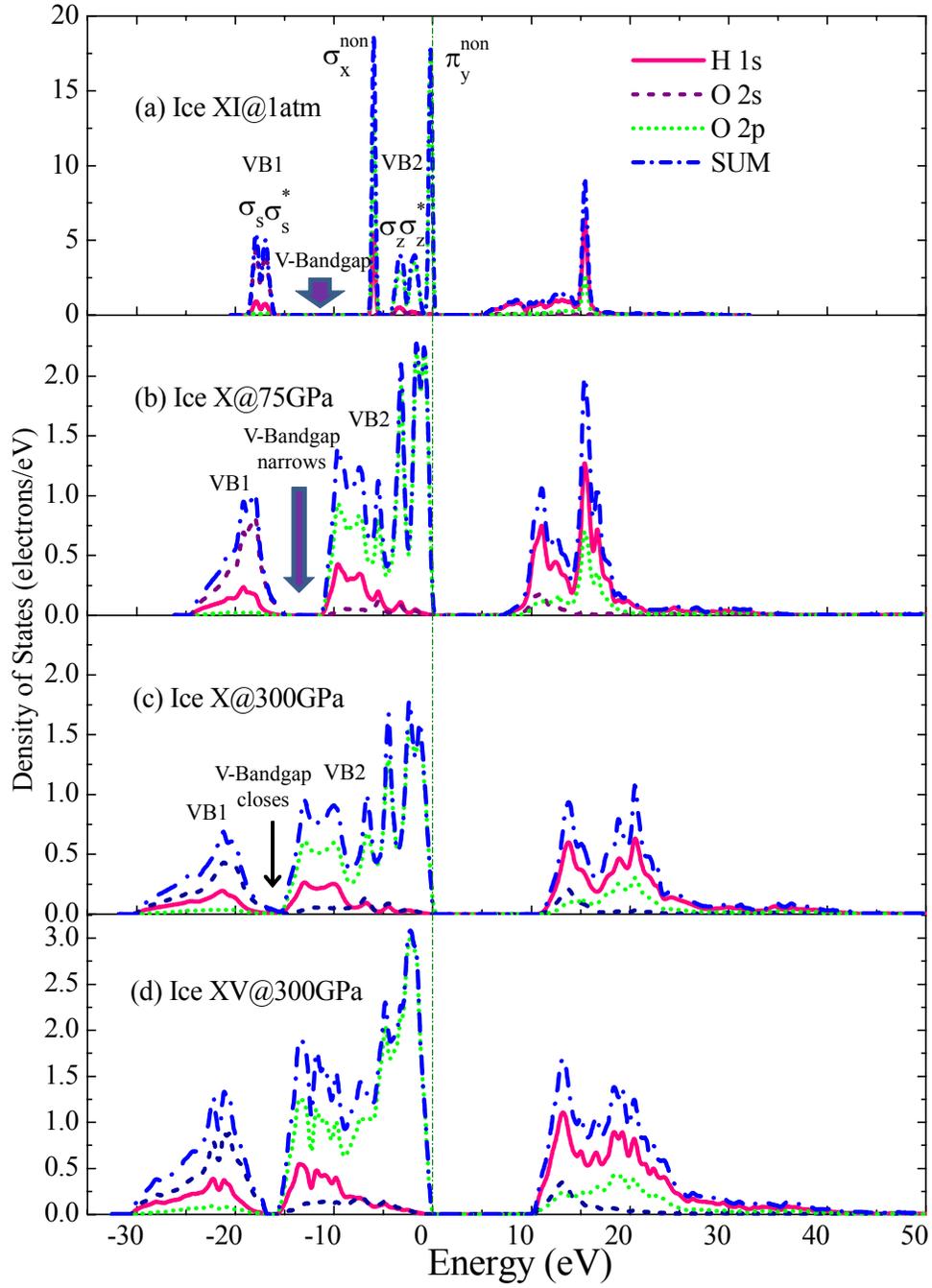

FIG. 9

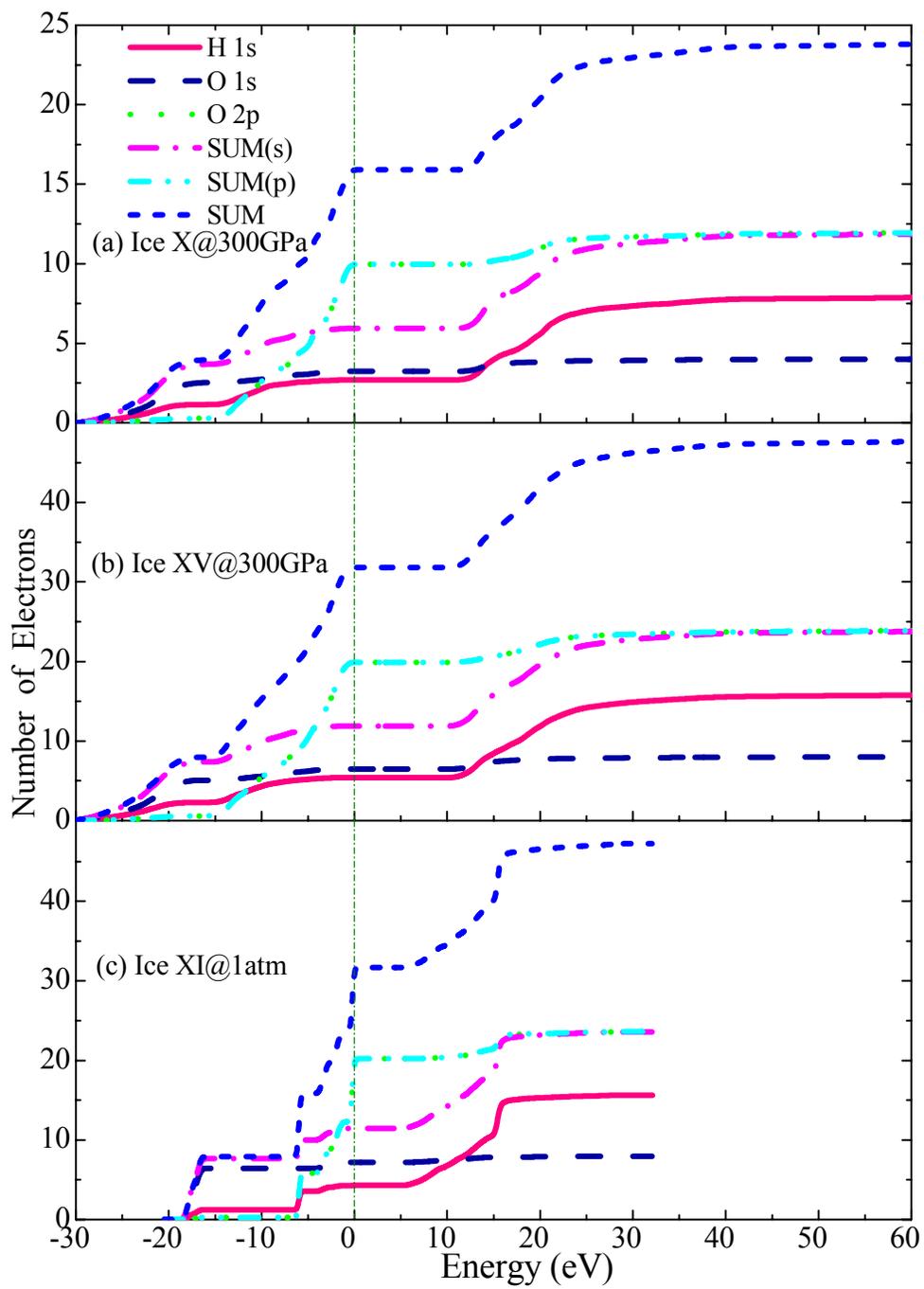

**FIG. 10**

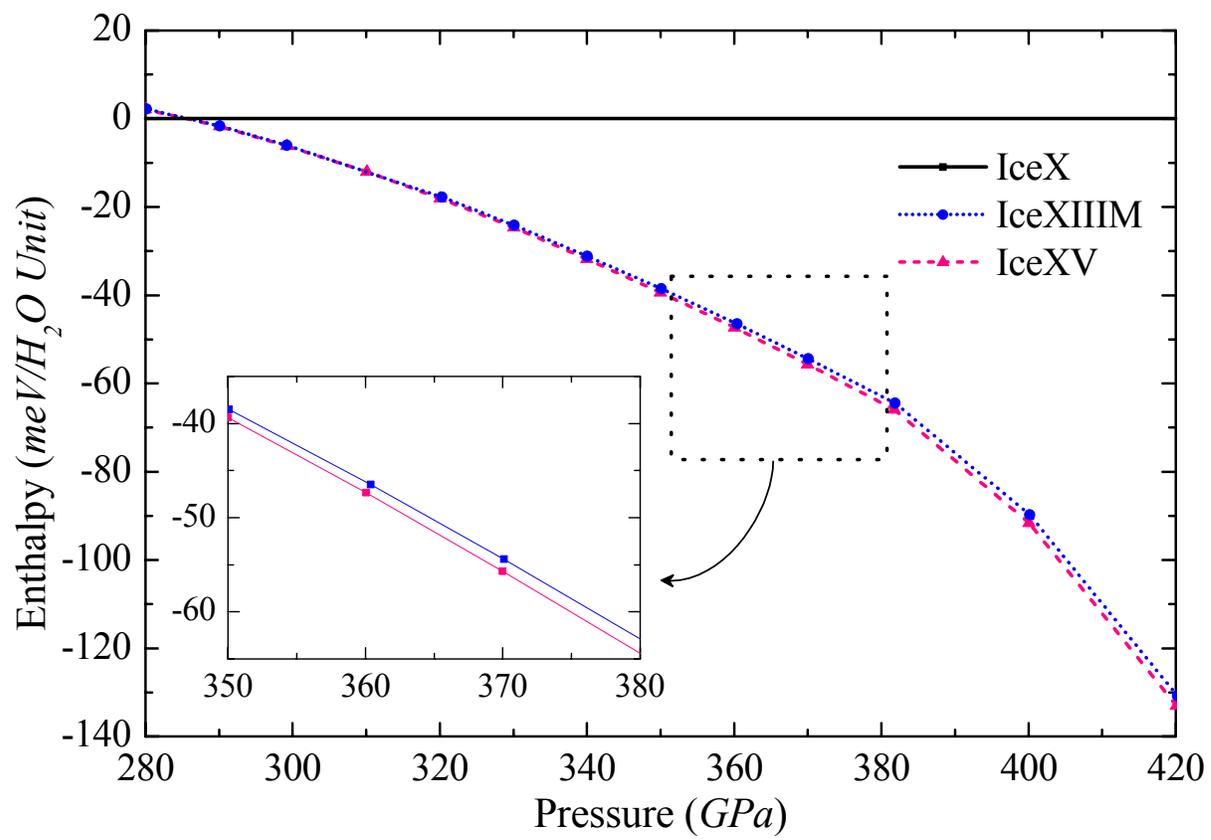

**FIG. 11**

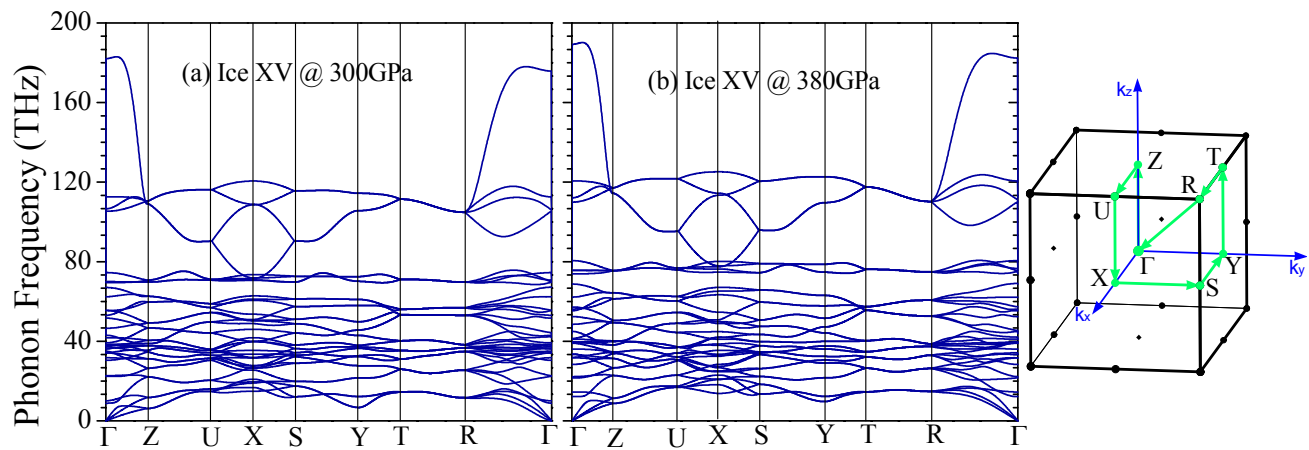

**FIG. 12**

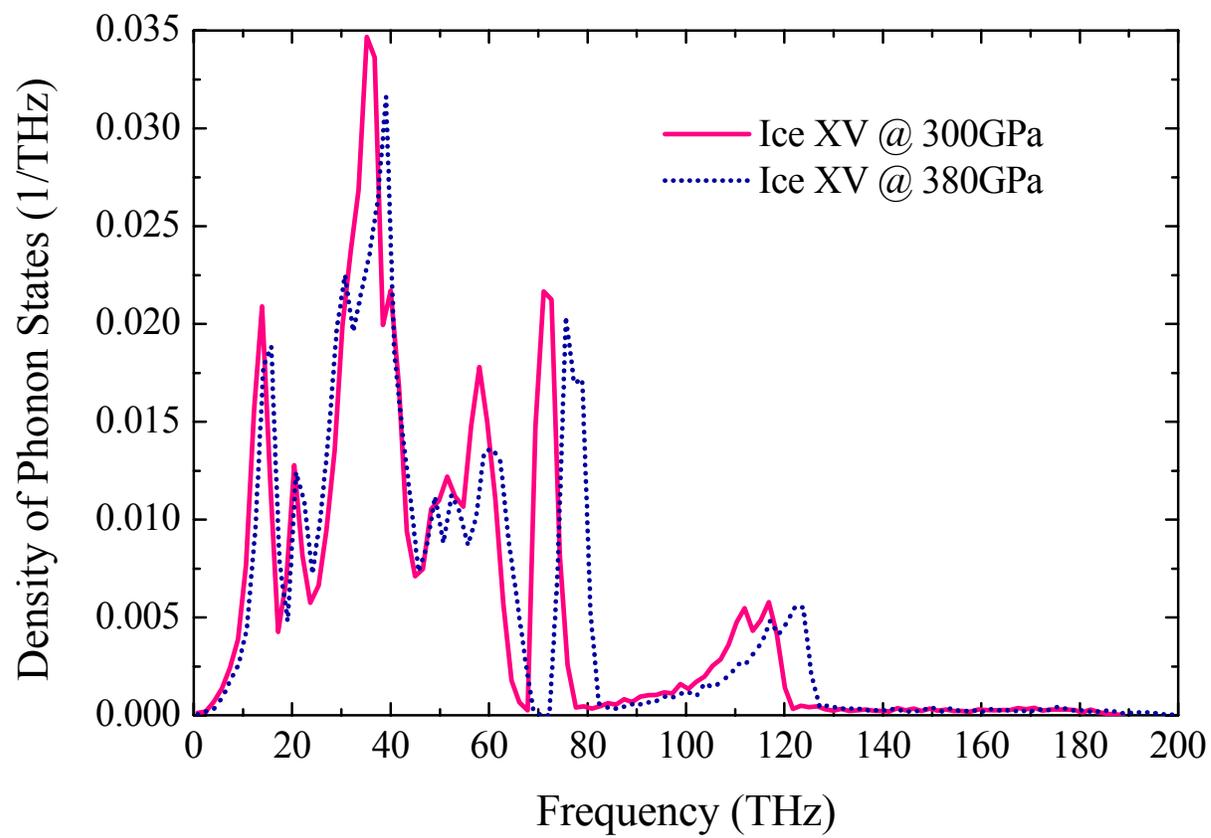

**FIG. 13**

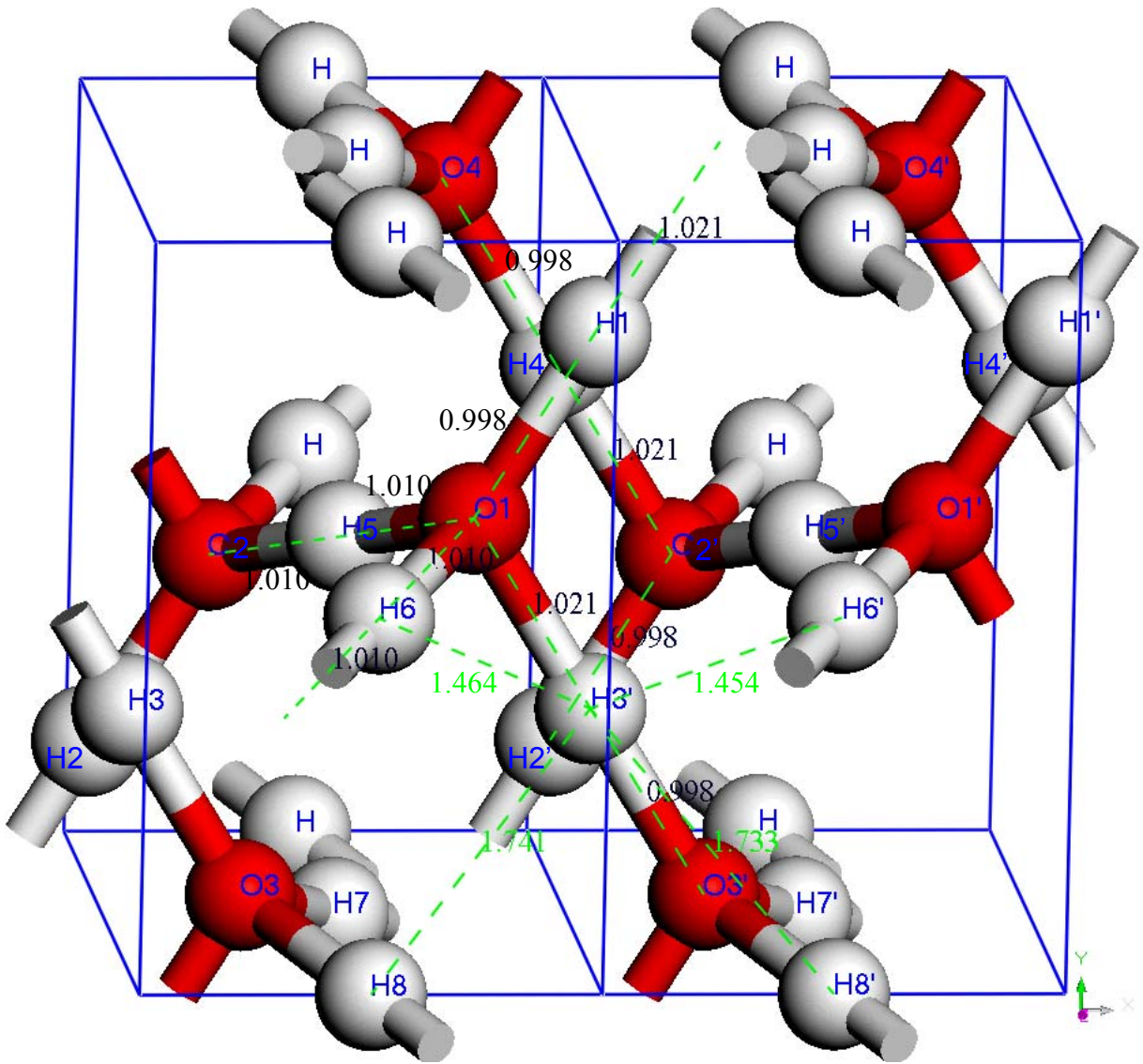

**FIG. 14**

# TABLE I

| Phase | Pressure (GPa) | a (Å) | b/a | c/a | Fractional coordinates H | O |
|---|---|---|---|---|---|---|
| XI | 1atm | 4.2394 | 1.7327 | 1.6320 | (0.6866,0.2318,0.0172) (0.0,0.3343,0.7987) (0.0,0.4606,0.9841) | (0.0,0.3323,0.9414) (0.5,0.1657,0.0662) |
|  | 1atm[a] | 4.4650 | 1.7601 | 1.6331 |  |  |
|  | 1atm[b] | 4.5019 | 1.7321 | 1.6278 | (0.6766,0.2252,0.0183) (0.0,0.3364,0.8037) (0.0,0.4637,0.9817) | (0.0,0.3352,0.9369) (0.5,0.8255,0.0631) |
| X | 62 | 2.6551 | 1 | 1 | (0.25,0.25,0.25) | (0.0,0.0,0.0) |
|  | 62[c] | 2.78 | 1 | 1 |  |  |
|  | 300 | 2.3583 | 1 | 1 | (0.25,0.25,0.25) | (0.0,0.0,0.0) |
|  | 380 | 2.2989 | 1 | 1 | (0.25,0.25,0.25) | (0.0,0.0,0.0) |
| XV | 300 | 2.1911 | 1.5817 | 1.5573 | (0.0020,0.1941,0.25) (0.5,0.5,0.0) | (0.2529,0.4437,0.25) |
|  | 380 | 2.0730 | 1.6553 | 1.6081 | (0.0033,0.1746,0.25) (0.5,0.5,0.0) | (0.2567,0.4218,0.25) |
| XIII[M] | 380[d] | 2.22 | 1.59 | 1.57 | (0.0,0.193,0.25) (0.5,0.5,0.0) | (0.252,0.443,0.25) |

[a] Experimental data from Ref. 24.

[b] Experimental data from Ref. 20.

[c] Experimental data from Ref. 16.

[d] Theoretical date from Ref. 1. Mind that the value of pressure could vary some while different pseudopotentials are used, especially within different *ab initio* programs.

# TABLE II

| Phase | Pressure (GPa) | Density (g/cm$^3$) | $Q_{tran}$ (|e|) | $Q_{eff}$ (e) | | |
|---|---|---|---|---|---|---|
| | | | | H 1s | O 2s | O 2p |
| XI | 1atm | 1.1107 | 0.46 | 0.54 | 1.80 | 5.11 |
| X | 75 | 3.3017 | 0.34 | 0.66 | 1.71 | 4.97 |
| | 300 | 4.5619 | 0.32 | 0.68 | 1.62 | 5.02 |
| | 380 | 4.9245 | 0.31 | 0.69 | 1.60 | 5.03 |
| XV | 300 | 4.6177 | 0.32 | 0.68 | 1.63 | 5.02 |
| | 380 | 5.0459 | 0.31 | 0.69 | 1.61 | 5.02 |

# TABLE III

| Phase | Pressure | Atom-Atom distance (Å) /Bond population | | |
|---|---|---|---|---|
| | | H-O | H-H | O-O |
| XI | 1atm | 0.9876/0.49 | 1.5817/-0.11 | 2.5954/-0.37 |
| | | 0.9877/0.49 | 1.5832/-0.10 | 2.5962/-0.12 |
| | | 0.9879/0.49 | 2.1493/-0.02 | |
| | | 1.6082/0.14 | 2.1549/-0.05 | |
| | | 1.6086/0.14 | 2.1572/-0.03 | |
| | | 1.6099/0.13 | 2.6469/-0.01 | |
| X | 75GPa | 1.1375/0.35 | 1.8575/-0.18 | 2.2750/-1.42 |
| | 300GPa | 1.0212/0.37 | 1.6675/-0.27 | 2.0423/-2.31 |
| | 380GPa | 0.9955/0.37 | 1.6256/-0.29 | 1.9909/-2.59 |
| XV | 300GPa | 1.0251/0.51 | 1.5397/-0.24 | 2.0434/-1.12 |
| | | 1.0290/0.49 | 1.7061/-0.06 | 2.0581/-1.09 |
| | | 1.0319/0.47 | 1.7329/-0.14 | 2.1712/-0.14 |
| | | 1.8401/-0.13 | 1.7329/-0.04 | |
| | | 1.8550/-0.19 | 1.7443/-0.09 | |
| | | 2.1907/-0.05 | 1.7496/-0.12 | |
| | | | 2.1727/0.02 | |
| | | | 2.4319/0.04 | |
| | 380GPa | 0.9977/0.55 | 1.4539/-0.31 | 1.9903/-1.21 |
| | | 1.0105/0.52 | 1.6669/-0.06 | 2.0210/-1.15 |
| | | 1.0213/0.46 | 1.7157/-0.15 | 2.0415/-0.22 |
| | | 1.7446/-0.15 | 1.7158/-0.02 | |
| | | 1.7787/-0.25 | 1.7326/-0.09 | |
| | | 2.2333/-0.04 | 1.7454/-0.12 | |
| | | | 2.0528/0.02 | |
| | | | 2.3921/0.06 | |

# TABLE IV

|        | Fractional coordinates of atoms |        |         |         |         |        | Distance (Å) |
|--------|--------|--------|---------|---------|---------|--------|--------------|
|        | H3     |        |         | O4      |         |        | H3-O4        |
| 300GPa | 1.0020 | 0.3059 | -0.2500 | -0.2529 | -0.0563 | 0.2500 | 2.1907       |
| 380GPa | 1.0009 | 0.3166 | -0.2500 | -0.2536 | -0.0698 | 0.2500 | 2.2333       |
| Δ      | -0.0011| 0.0107 | 0.0000  | -0.0007 | -0.0135 | 0.0000 | 0.0426       |

[1] M. Benoit, M. Bernasconi, P. Focher, and M. Parrinello, Phys. Rev. Lett. **76**, 2934 (1996).

[2] M. Bernasconi, M. Benoit, M. Parrinello, G. L. Chiarotti, P. Focher, and E. Tosatti, Phys. Scr. **T66**, 98 (1996).

[3] Kanani K. M. Leea, L. Robin Benedetti, Raymond Jeanloz et al., J. Chem. Phys. **125**, 014701 (2006).

[4] P. V. Hobbs, *Ice Physics* (Clarendon, Oxford, 1974).

[5] W. B. Hubbard, *Planetary Interiors* (Van Norstrand Reinhold, New York, 1984).

[6] W. J. Nellis et al., Science **240**, 779 (1988).

[7] O Mishima and HE. Stanley, Nature **396**, 329 (1998).

[8] I. Ohmine and S. Saito, Acc. Chem. Res. **32**, 741 (1999).

[9] R. J. Hemley, Annu. Rev. Phys. Chem. **51**, 763 (2000).

[10] T. Loerting and N. Giovambattista, J. Phys.: Condens. Matter **18**, 919 (2006).

[11] John S. Tse, Dennis D. Klug, Malcolm Guthrie, Chris A. Tulk, Chris J. Benmore, and Jacob Urquidi, Phys. Rev. B **71**, 214107 (2005).

[12] V. F. Petrenko, and R. W. Whitworth, *Physics of Ice* (Oxford University Press, Oxford, 1999).

[13] C. Cavazzoni, G.L. Chiarotti, S. Scandolo, E. Tosatti, M. Bernasconi, and M. Parrinello, Science **283**, 44 (1999).

[14] Theoretical work: M. Benoit, D. Marx, and M. Parrinello, Nature **392**, 258 (1998); J. Teixeira, Nature **392**, 232 (1998); M. Benoit, A. H. Romero, and D. Marx, Phys. Rev. Lett. **89**, 145501 (2002).

[15] Experimental work: Ph. Pruzan and M. Gauthier, J. Mol. Struct. **177**, 429 (1988); J.D. Londono, W.F. Kuhs, and J.L. Finney, J. Chem. Phys. **98**, 4878 (1993); P. Loubeyre, R. LeToullec, E. Wolanin, M. Hanfland, and D. Hausermann, Nature **397**, 503 (1999); A. F. Goncharov, V. V. Struzhkin, M. S. Somayazulu, R. J. Hemley, and H. K. Mao, Science **273**, 218 (1996); P. Pruzan, J. C. Chervin, and B. Canny, J. Chem. Phys. **99**, 9842 (1993); A. F. Goncharov, V. V. Struzhkin, H. K. Mao, and R. J. Hemley, Phys. Rev. Lett. **83**, 1998 (1999); M. Song, H. Yamawaki, H. Fujihisa, M. Sakashita, and K. Aoki, Phys. Rev. B **68**, 014106 (2003); Ph. Pruzan, J. C. Chervin, E. Wolanin, B. Canny, M. Gauthier, and M. Hanfland, J. Raman Spectrosc. **34**, 591 (2003).

[16] R. J. Hemley, A. P. Jephcoat, and H. K. Mao et al., Nature **330**, 737 (1987).

[17] A. F. Goncharov, V. V. Struzkhin, M. S. Somayazulu, R. J. Hemley, and H. K. Mao, Science **273**, 218 (1996).

[18] C. G. Salzmann, P. G. Radaelli, A. Hallbrucker, E. Mayer, and J. L. Finney, Science **311**, 1758 (2006).

[19] T. Matsuo, Y. Tajima, and H. Suga, J. Phys. Chem. Solids **47**, 165 (1986).

[20] A. J. Leadbetter, R. C. Ward, J. W. Clark, P. A. Tucker, T. Matsuo, and H. Suga, J. Chem. Phys. **82**, 424 (1985).

[21] R. Howe and R. W. Whitworth, J. Chem. Phys. **90**, 4550 (1989).

[22] S. M. Jackson and R. W. Whitworth, J. Chem. Phys. **103**, 7647 (1995).

[23] S. M. Jackson, J. Phys. Chem. B **101**, 6177 (1997).

[24] C. M. B. Line and R. W. Whitworth, J. Chem. Phys. **104**, 10008 (1996).

[25] S. M. Jackson, V. M. Nield, R. W. Whitworth, M. Oguro, and C. C. Wilson, J. Phys. Chem. B **101**, 6142 (1997).

[26] H. Fukazawa, S. Ikeda, M. Oguro, T. Fukumura, and S. Mae, J. Phys. Chem. B **106**, 6021 (2002).

[27] Ph. Pruzan, J. Mol. Struct. **322**, 279 (1994).

[28] F. H. Stillinger, J. Phys. Chem. **87**, 4281 (1983).

[29] C. Lee, D. Vanderbilt, K. Laasonen, R. Car, and M. Parrinello, Phys. Rev. Lett. **69**, 462 (1992); Phys. Rev. B **47**, 4863 (1993).

[30] B. Kamb and B. L. Davis, Proc. Natl. Acad. Sci. U.S.A. **52**, 1433 (1964).

[31] W. B. Holzapfel, J. Chem. Phys. **56**, 712 (1972).

[32] M. D. Segall, P. J. D. Lindan, M. J. Probert, C. J. Pickard, P. J. Hasnip, S. J. Clark, and M. C. Payne, J. Phys.: Cond. Matt. **14(11)**, 2717 (2002).

[33] J.P. Perdew, K. Burke, and M. Ernzerhof, Phys. Rev. Lett. **77**, 3865 (1996).

[34] H. J. Monkhorst and J. D. Pack, Phys. Rev. B **13**, 5188 (1976).

[35] G. J. Ackland, Phys. Rev. Lett. **80**, 2233 (1998).

[36] Sanchez Portal, E. Artacho, and J. M. Soler, Solid State Commun. **95**, 685 (1995); M. D. Segall, R. Shah, C. J. Pickard, and M. C. Payne, Phys. Rev. B **54**, 16317 (1996).

[37] K. Parlinski, Z. Q. Li, and Y. Kawazoe, Phys. Rev. Lett. **78**, 4063 (1997); K. Parlinski, Software PHONON, Cracow, 2005.

[38] J. S. Tse, G. Frapper, A. Ker, R. Rousseau, and D. D. Klug, Phys. Rev. Lett. **82**, 4472 (1999).

[39] Jian Sun and Hui-Tian Wang et al., Phys. Rev. B **71**, 125132 (2005).

[40] Jian Sun et al., Phys. Rev. B **73**, 045108 (2006).

[41] R. Car and M. Parrinello, Phys. Rev. Lett. **55**, 2471 (1985).

[42] D. Marx and J. Hutter, in *Modern Methods and Algorithms of Quantum Chemistry*, edited by J. Grotendorst (NIC, FZ Jülich, 2000), Vol. **1**, p.301.

[43] Ph. Pruzan, E. Wolanin, M. Gauthier, and et al., J. Phys. Chem. B **101**, 6230 (1997).

[44] K. Aoki, H. Yamawaki, M. Sakashita, and H. Fujihisa, Phys. Rev. B **54**, 15673 (1996).

[45] B. Hammer, L. B. Hansen, and J. K. Norskov, Phys. Rev. B **59**, 7413 (1999).

[46] J. P. Perdew, J. A. Chevary, and S. H. Vosko et al., Phys. Rev. B **46**, 6671 (1992).

[47] D. M. Ceperley and B. J. Alder, Phys. Rev. Lett. **45**, 566 (1980).

[48] P.E. Blöchl, Phys. Rev. B **50**, 17953 (1994); G. Kresse and J. Joubert, Phys. Rev. B **59**, 1758 (1999).

[49] C. Lobban, J. L. Finney, and W. F. Kuhs, Nature **391**, 268 (1998).

[50] W. F. Kuhs and T. C. Hansen, Rev. Mineral Geochem. **63**, 171 (2006).